# Continuous Time Dynamic Topic Models

**Chong Wang**
Computer Science Dept.
Princeton University
Princeton, NJ 08540

**David Blei**
Computer Science Dept.
Princeton University
Princeton, NJ 08540

**David Heckerman**
Microsoft Research
One Microsoft Way
Redmond, WA 98052

## Abstract

In this paper, we develop the continuous time dynamic topic model (cDTM). The cDTM is a dynamic topic model that uses Brownian motion to model the latent topics through a sequential collection of documents, where a "topic" is a pattern of word use that we expect to evolve over the course of the collection. We derive an efficient variational approximate inference algorithm that takes advantage of the sparsity of observations in text, a property that lets us easily handle many time points. In contrast to the cDTM, the original discrete-time dynamic topic model (dDTM) requires that time be discretized. Moreover, the complexity of variational inference for the dDTM grows quickly as time granularity increases, a drawback which limits fine-grained discretization. We demonstrate the cDTM on two news corpora, reporting both predictive perplexity and the novel task of time stamp prediction.

## 1 Introduction

Tools for analyzing and managing large collections of electronic documents are becoming increasingly important. In recent years, *topic models*, which are hierarchical Bayesian models of discrete data, have become a widely used approach for exploratory and predictive analysis of text. Topic models, such as latent Dirichlet allocation (LDA) and the more general discrete component analysis [3, 4], posit that a small number of distributions over words, called topics, can be used to explain the observed collection. LDA is a probabilistic extension of latent semantic indexing (LSI) [5] and probabilistic latent semantic indexing (pLSI) [11]. Owing to its formal generative semantics, LDA has been extended and applied to authorship [19], email [15], computer vision [7], bioinformatics [18], and information retrieval [24]. For a good review, see [8].

Most topic models assume the documents are exchangeable in the collection, i.e., that their probability is invariant to permutation. Many document collections, such as news or scientific journals, evolve over time. In this paper, we develop the *continuous time dynamic topic model* (cDTM), which is an extension of the discrete dynamic topic model (dDTM) [2]. Given a sequence of documents, we infer the latent topics and how they change through the course of the collection.

The dDTM uses a state space model on the natural parameters of the multinomial distributions that represent the topics. This requires that time be discretized into several periods, and within each period LDA is used to model its documents. In [2], the authors analyze the journal *Science* from 1880-2002, assuming that articles are exchangeable within each year. While the dDTM is a powerful model, the choice of discretization affects the memory requirements and computational complexity of posterior inference. This largely determines the resolution at which to fit the model.

To resolve the problem of discretization, we consider time to be continuous. The continuous time dynamic topic model (cDTM) proposed here replaces the discrete state space model of the dDTM with its continuous generalization, Brownian motion [14]. The cDTM generalizes the dDTM in that the only discretization it models is the resolution at which the time stamps of the documents are measured.

The cDTM model will, generally, introduce many more latent variables than the dDTM. However, this seemingly more complicated model is simpler and more efficient to fit. As we will see below, from this formulation the variational posterior inference procedure can take advantage of the natural sparsity of text, the fact that not all vocabulary words are used at each measured time step. In fact, as the resolution gets finer, fewer and fewer words are used.

This provides an inferential speed-up that makes it possible to fit models at varying granularities. As examples, journal articles might be exchangeable within an issue, an assumption which is more realistic than one where they are exchangeable by year. Other data, such as news, might experience periods of time without any observation. While the dDTM requires representing all topics for the discrete ticks within these periods, the cDTM can analyze such data without a sacrifice of memory or speed. With the cDTM, the granularity can be chosen to maximize model fitness rather than to limit computational complexity.

We note that the cDTM and dDTM are not the only topic models to take time into consideration. Topics over time models (TOT) [23] and dynamic mixture models (DMM) [25] also include timestamps in the analysis of documents. The TOT model treats the time stamps as observations of the latent topics, while DMM assumes that the topic mixture proportions of each document is dependent on previous topic mixture proportions. In both TOT and DMM, the topics themselves are *constant*, and the time information is used to better discover them. In the setting here, we are interested in inferring evolving topics.

The rest of the paper is organized as follows. In section 2 we describe the dDTM and develop the cDTM in detail. Section 3 presents an efficient posterior inference algorithm for the cDTM based on sparse variational methods. In section 4, we present experimental results on two news corpora.

## 2 Continuous time dynamic topic models

In a time stamped document collection, we would like to model its latent topics as changing through the course of the collection. In news data, for example, a single topic will change as the stories associated with it develop. The discrete-time dynamic topic model (dDTM) builds on the exchangeable topic model to provide such machinery [2]. In the dDTM, documents are divided into sequential groups, and the topics of each slice evolve from the topics of the previous slice. Documents in a group are assumed exchangeable.

More specifically, a topic is represented as a distribution over the fixed vocabulary of the collection. The dDTM assumes that a discrete-time state space model governs the evolution of the natural parameters of the multinomial distributions that represent the topics. (Recall that the natural parameters of the multinomial are the logs of the probabilities of each item.) This is a time-series extension to the logistic normal distribution [26].

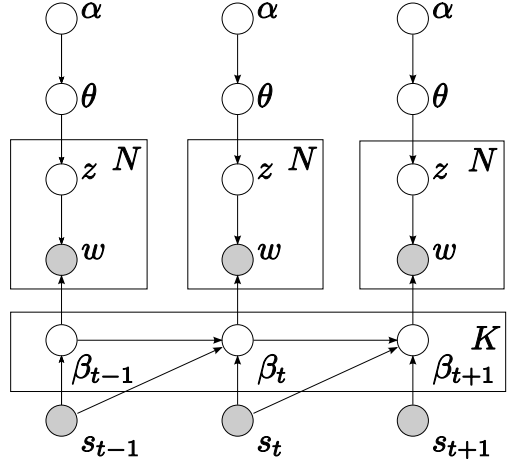

Figure 1: Graphical model representation of the cDTM. The evolution of the topic parameters $\beta_t$ is governed by Brownian motion. The variable $s_t$ is the observed time stamp of document $d_t$.

A drawback of the dDTM is that time is discretized. If the resolution is chosen to be too coarse, then the assumption that documents within a time step are exchangeable will not be true. If the resolution is too fine, then the number of variational parameters will explode as more time points are added. Choosing the discretization should be a decision based on assumptions about the data. However, the computational concerns might prevent analysis at the appropriate time scale.

Thus, we develop the continuous time dynamic topic model (cDTM) for modeling sequential time-series data with arbitrary granularity. The cDTM can be seen as a natural limit of the dDTM at its finest possible resolution, the resolution at which the document time stamps are measured.

In the cDTM, we still represent topics in their natural parameterization, but we use Brownian motion [14] to model their evolution through time. Let $i$, $j$ ($j > i > 0$) be two arbitrary time *indexes*, $s_i$ and $s_j$ be the time stamps, and $\Delta_{s_j,s_i}$ be the elapsed time between them. In a $K$-topic cDTM model, the distribution of the $k^{th}$ ($1 \le k \le K$) topic's parameter at term $w$ is:

$$\begin{aligned}\beta_{0,k,w} &\sim \mathcal{N}(m, v_0) \\ \beta_{j,k,w} | \beta_{i,k,w}, s &\sim \mathcal{N}\left(\beta_{i,k,w}, v\Delta_{s_j,s_i}\right),\end{aligned} \quad (1)$$

where the variance increases linearly with the lag.

This construction is used as a component in the full generative process. (Note: if $j = i+1$, we write $\Delta_{s_j,s_i}$ as $\Delta_{s_j}$ for short.)

1. For each topic $k, 1 \le k \le K$,

    (a) Draw $\beta_{0,k} \sim \mathcal{N}(m, v_0 I)$.

2. For document $d_t$ at time $s_t$ $(t > 0)$:
   (a) For each topic $k, 1 \leq k \leq K$,
      i. From the Brownian motion model, draw
         $\beta_{t,k}|\beta_{t-1,k}, s \sim \mathcal{N}(\beta_{t-1,k}, v\Delta_{s_t} I)$.
   (b) Draw $\theta_t \sim \text{Dir}(\alpha)$.
   (c) For each word,
      i. Draw $z_{t,n} \sim \text{Mult}(\theta_t)$.
      ii. Draw $w_{t,n} \sim \text{Mult}(\pi(\beta_{t,z_{t,n}}))$.

The function $\pi$ maps the multinomial natural parameters, which are unconstrained, to its mean parameters, which are on the simplex,

$$\pi(\beta_{t,k})_w = \frac{\exp(\beta_{t,k,w})}{\sum_w \exp(\beta_{t,k,w})}. \qquad (2)$$

The cDTM is illustrated in Figure 1.

The cDTM can be seen as a generalization of the dDTM. Both models assume that the log probability of a term exhibits variance over an interval of time between observations. In the dDTM, this interval is evenly divided into discrete ticks. A parameter controls the variance at each tick, and the variance across the whole interval is that parameter multiplied by the number of ticks. As a consequence of this representation, the topic, i.e., the full distribution over terms, is explicitly represented at each tick. For fine-grained time series, this leads to high memory requirements for posterior inference, even if the observations are sparsely distributed throughout the timeline.

In the cDTM, however, the variance is a function of the lag between observations, and the probabilities at discrete steps between those observations need not be considered. Inference, as we will see below, can be handled sparsely. Thus, choosing the right granularity becomes a modeling issue rather than one governed by computational concerns. A dDTM is obtained with a cDTM by measuring the time stamps of the documents at the desired granularity.

Akin to Brownian motion as the limiting process of a discrete-time Gaussian random walk [6], the cDTM is the limiting process of the dDTM. Denote the per-tick variance in the dDTM by $\sigma^2$, and note that it is a function of the tick granularity (to make models comparable). The cDTM is the limiting model in this setting as $\sigma^2$ approaches zero. We emphasize that with the cDTM, we need not represent the log probabilities at the ticks between observed documents. This perspective is illustrated in Figure 2.

## 3 Sparse variational inference

The central problem in topic modeling is posterior inference, i.e., determining the distribution of the la-tent topic structure conditioned on the observed documents. In sequential topic models, this structure comprises the per-document topic proportions $\theta_d$, per-word topic assignments $z_{d,n}$, and the $K$ sequences of topic distributions $\beta_{t,k}$. The true posterior is not tractable [2]. We must appeal to an approximation.

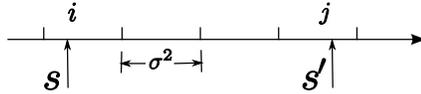

Figure 2: Documents are available only at time $s$ and $s'$, and no documents between them. When $\sigma^2 \to 0$, the dDTM becomes a cDTM, and we no longer need to represent the steps between $i$ and $j$.

Several approximate inference methods have been developed for topic models. The most widely used are variational inference [3, 20] and collapsed Gibbs sampling [9]. In the sequential setting collapsed Gibbs sampling is not an option because the distribution of words for each topic is not conjugate to the word probabilities. Thus, we employed variational methods.

The main idea behind variational methods is to posit a simple family of distributions over the latent variables, indexed by free *variational parameters*, and to find the member of that family which is closest in Kullback-Leibler divergence to the true posterior. Good overviews of this methodology can be found in [12] and [22]. For continuous time processes, variational inference has been applied in Markov jump processes [1] and diffusion processes [17], where the variational distributions are also random processes.

For the cDTM described above, we adapt variational Kalman filtering [2] to the continuous time setting. For simplicity, assume that one document occurs at each time point. In their algorithm, the variational distribution over the latent variables is:

$$q(\beta_{1:T}, z_{1:T,1:N}, \theta_{1:T} \,|\, \hat{\beta}, \phi, \gamma) = \\ \prod_{k=1}^{K} q(\beta_{1,k}, \ldots, \beta_{T,k}|\hat{\beta}_{1,k}, \ldots, \hat{\beta}_{T,k}) \times \\ \prod_{t=1}^{T} \left( q(\theta_t|\gamma_t) \prod_{n=1}^{N_t} q(z_{t,n}|\phi_{t,n}) \right). \qquad (3)$$

The variational parameters are a Dirichlet $\gamma_t$ for the per-document topic proportions, multinomials $\phi$ for each word's topic assignment, and $\hat{\beta}$ variables, which are "observations" to a variational Kalman filter.

These variables are fit such that the approximate posterior is close to the true posterior. From the variational Kalman filter, the $\beta_{k,t}, 1 \leq t \leq T$ retain their chained structure in the variational distribution. Vari-ational inference proceeds by coordinate ascent, updating each of these parameters to minimize the KL between the true posterior and variational posterior.

For simplicity, now we consider a model with only *one* topic. These calculations are simpler versions of those we need for the more general latent variable model but exhibit the essential features of the algorithm. For the cDTM, we assume a similar variational distribution, with the same variational Dirichlet and variational multinomials for the per-document variables. The cDTM updates for these parameters are identical to those in [2], and we do not replicate them here.

In principle, we can directly use the variational Kalman filtering algorithm for the cDTM by replacing the state space model with Brownian motion. Let $\mathcal{V}$ be the size of the vocabulary. While conceptually straightforward, this will yield $\mathcal{V}T$ variational parameters in the vectors $\hat{\beta}_{1:T}$. When $T$ and $\mathcal{V}$ are large, as in a fine-graned model, posterior inference will require massive amounts of time and memory. Thus, we develop a sparse variational inference procedure, which significantly improves its complexity without sacrificing accuracy.

The main idea behind the sparse variational Kalman filtering algorithm is that if certain $\beta_{t,w}$ do not describe any term emissions, i.e., there are no observations of $w$ at $t$, then the true posterior of $\beta_{t,w}$ is only determined by the observations of the other words at that time. Therefore, we don't need to explicitly represent $\hat{\beta}_{t,w}$ for those $w$ that are not observed.

Figure 3 illustrates the idea behind sparse variational inference for the cDTM. In Figure 3, the variational posterior of the log probability of a word $\beta_{t,w}$ is determined by the variational observations of the observed words. From the belief propagation point of view, the belief propagated from $\beta_{t,w}$ to node $\beta_{t+2,w}$ is not revised by term $w$, and this property is retained in the sparse variational inference algorithm. The probability of variational observation $\hat{\beta}_{t,w}$ given $\beta_{t,w}$ is a Gaussian:

$$\hat{\beta}_{t,w}|\beta_{t,w} \sim \mathcal{N}(\beta_{t,w}, \hat{v}_t). \quad (4)$$

We next describe the forward-backward algorithm for the sparse variational Kalman filter, which is needed to compute the expectations for updating the variational parameters. For a certain term $w$, the variational forward distribution $p(\beta_{t,w}|\hat{\beta}_{i,i\leq t,w})$ is a Gaussian [13] and can be characterized as follows.

$$\begin{aligned}
\beta_{t,w}|\hat{\beta}_{i,i\leq t,w} &\sim \mathcal{N}(m_{t,w}, V_{t,w}) \\
m_{t,w} &= \mathbb{E}(\beta_{t,w}|\hat{\beta}_{i,i\leq t,w}) \\
V_{t,w} &= \mathbb{E}((\beta_{t,w}-m_{t,w})^2|\hat{\beta}_{i,i\leq t,w}). \quad (5)
\end{aligned}$$

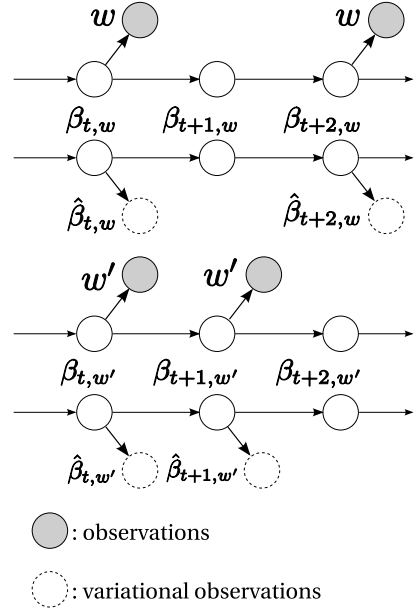

Figure 3: A simplified graphical model shows how sparse variational inference works with only single topic. Note this generation process needs normalization to $\beta_t$ according to Equation 2, but this will not affect the sparse solution. For term $w$, there are no observations at time index $t+1$ (or time $s_{t+1}$), the corresponding variational observations don't appear at time index $t + 1$. For term $w'$, there are no observations at time index $t+2$ (or time $s_{t+2}$), the corresponding variational observations don't appear at time index $t + 2$.

If $w$ is not observed at time step $t$ then

$$\begin{aligned}
m_{t,w} &= m_{t-1,w} \\
V_{t,w} &= P_{t,w}, \\
P_{t,w} &= V_{t-1,w} + v\Delta_{s_t}, \quad (6)
\end{aligned}$$

which means that the forward mean remains the same as the previous step. Otherwise,

$$\begin{aligned}
m_{t,w} &= \frac{\hat{\beta}_{t,w}P_{t,w} + \hat{v}_t m_{t-1,w}}{P_{t,w} + \hat{v}_t} \\
V_{t,w} &= \hat{v}_t \frac{P_{t,w}}{P_{t,w} + \hat{v}_t} \\
\hat{\beta}_{t,w}|\hat{\beta}_{i,i\leq t-1,w} &\sim \mathcal{N}(m_{t-1,w}, P_{t,w} + \hat{v}_t). \quad (7)
\end{aligned}$$

Similarly, the variational backward distribution $p(\beta_{t,w}|\hat{\beta}_{i,i\leq T,w})$ is also a Gaussian:

$$\begin{aligned}
\beta_{t,w}|\hat{\beta}_{i,i\leq T,w} &\sim \mathcal{N}(\widetilde{m}_{t,w}, \widetilde{V}_{t,w}) \\
\widetilde{m}_{t,w} &= \mathbb{E}(\beta_{t,w}|\hat{\beta}_{i,i\leq T,w}) \\
\widetilde{V}_{t,w} &= \mathbb{E}((\beta_{t,w}-\widetilde{m}_{t,w})^2|\hat{\beta}_{i,i\leq T,w}). \\
\widetilde{m}_{t-1,w} &= m_{t-1,w}\frac{f_t v}{P_{t,w}} + \widetilde{m}_{t,w}\frac{V_{t-1,w}}{P_{t,w}}
\end{aligned}$$

$$\widetilde{V}_{t-1,w} = V_{t-1,w} + \frac{V_{t-1,w}^2}{P_{t,w}^2}(\widetilde{V}_{t,w} - P_{t,w}). \quad (8)$$

With this forward-backward computation in hand, we turn to optimizing the variational observations $\hat{\beta}_{w,k}$ in the sparse setting. Equivalent to minimizing KL is tightening the bound on the likelihood of the observations given by Jensen's inequality [12].

$$\mathcal{L}(\hat{\beta}) \geq \sum_{t=1}^{T} \mathbb{E}_q\left[(\log p(\mathbf{w}_t|\beta_t) + \log p(\beta_t|\beta_{t-1})\right] + H(q), \quad (9)$$

where $H(q)$ is the entropy. This is simplified to

$$\mathcal{L}(\hat{\beta}) \geq \sum_{t=1}^{T} \mathbb{E}_q\left[\log p(\mathbf{w}_t|\beta_t) - \log q(\hat{\beta}_t|\beta_t)\right] \\ + \sum_{t=1}^{T} \log q(\hat{\beta}_t|\hat{\beta}_{i,i\leq t-1}), \quad (10)$$

We use $\delta_{t,w} = 1$ or $0$ to represent whether $\hat{\beta}_{t,w}$ is in the variational observations or not. Then the terms above are

$$\mathbb{E}_q \log q(\mathbf{w}_t|\beta_t) \geq \sum_w n_{t,w} \widetilde{m}_{t,w} \\ - n_t \log \sum_w \exp(\widetilde{m}_{t,w} + \widetilde{V}_{t,w}/2)$$

$$\mathbb{E}_q \log p(\hat{\beta}_t|\beta_t) = \sum_w \delta_{t,w} \mathbb{E}_q \log q(\hat{\beta}_{t,w}|\beta_{t,w})$$

$$\log q(\hat{\beta}_t|\hat{\beta}_{i,i\leq t-1}) = \sum_w \delta_{t,w} \log q(\hat{\beta}_{t,w}|\hat{\beta}_{i,i\leq t-1,w}).$$

The count of $w$ in document $d_t$ is $n_{t,w}$ and $n_t = \sum_w n_{t,w}$.

Thus, to optimize the variational observations, we need only to compute the derivative $\partial \mathcal{L}/\partial \hat{\beta}_{t,w}$ for those $\delta_{t,w} = 1$. The general memory requirement is $\mathcal{O}(\sum_t \sum_w \delta_{t,w})$—the sum of the number of unique terms at each time point—which is usually much smaller than $\mathcal{O}(\mathcal{V}T)$, the memory requirement for the densely represented algorithm. Formally, we can define the sparsity of the data set to be

$$\text{sparsity} = 1 - \left(\sum_t \sum_w \delta_{t,w}\right)/(\mathcal{V}T), \quad (11)$$

which we will compute for several data sets in the next section. Finally, we note that we use the conjugate gradient algorithm [16] to optimize the variational observations from these partial derivatives.

As an example of the speed-up offered by sparse variational inference, consider the *Science* corpus from 1880-2002, analyzed by [2], which contains 6243 issues of the magazine. Note that these issues are not evenly spaced over the time line. In the dDTM, these documents were separated by years. To analyze them at a finer scale, e.g., issue by issue, one needs to consider 6243 time points. With a vocabulary size of 5000, for a 10-topic setting, the cDTM requires 0.8G memory while the dDTM requires 2.3G memory, nearly 3 times larger. The sparsity of *Science* is 0.65. This means that a term only appears in about a third of the total time points.

| Data set | Sparsity | | | |
|---|---|---|---|---|
| | Hour | Day | Week | Month |
| AP | 0.93 | 0.68 | 0.12 | – |
| Election 08 | – | 0.95 | 0.79 | 0.50 |

Table 1: Sparsity for two data sets where available. Higher numbers indicate a sparser data set and more efficiency for the cDTM over the dDTM.

## 4 Experiments

In this section, we demonstrate the cDTM model on two news corpora. We report predictive perplexity and a results on the novel task of time stamp prediction.

### 4.1 News Corpora

We used two news corpora. First, "AP" is a subset from the TREC AP corpus [10] containing the news from 05/01/1988 to 06/30/1988. We extracted the documents about the presidential election in 1988 resulting in $1,342$ documents. These documents are time stamped by hour. Second, the "Election 08" data are summaries of the top articles from Digg[1] classified as being part of the 2008 presidential election. We used articles from 02-27-2007 to 02-22-2008. This data set has $1,040$ summaries. Time is measured in days.

Table 1 shows the sparsity information for these data in terms of the resolution at which we can analyze them. This illustrates the gain in efficiency of the cDTM. For example, in the day setting of the Election 08 data, the sparsity is 0.95. The dDTM model will need at least 20 times more parameters than the cDTM to analyze the data at this resolution.

### 4.2 Per-Word Predictive Perplexity

Let $D_t$ be the set of documents at time index $t$. We performed approximate posterior inference on these data with the cDTM at different levels of granularity. To make models comparable, we set the variance across the entire period to be the same (see Equation 1). We evaluated the models with perplexity. Specifi-cally, we computed the per-word predictive perplexity of the documents at time $t$ based on the data of the previous $t-1$ time indices,

$$\text{perplexity}_{\text{pw}}(t) = \exp\left\{-\frac{1}{|D_t|}\sum_{d \in D_t}\frac{\log p(\mathbf{w}_d|D_{1:t-1})}{N_d}\right\}. \quad (12)$$

---
[1] http://digg.com

Note that lower numbers are better.

Since each document is predicted exactly once in all models at different granularities, we also compute the *averaged* per-word perplexity over the time line, which is defined as

$$\text{perplexity}_{\text{pw}} = \exp\left\{-\frac{\sum_{d \in D}\log p(\mathbf{w}_d)}{\sum_{d \in D} N_d}\right\}. \quad (13)$$

In the AP data, we made predictions from 5/15/1988 to 05/29/1988. In the Election 08 data, we made predictions from 04/26/2007 to 02/22/2008. Figure 4 shows the results of the per-word predictive perplexity over the time line on both data sets for the 10 topic model. Figure 5 shows the results of average per-word perplexity for 1, 3, 5 and 10 topics.

From the computational perspective, we note that the sparse inference algorithm lets us fit models of different granularities efficiently. For the AP data, the day model and week are almost comparable. Models with 5 and 10 topics perform better.

In the Election 08 data, the 1-topic model performs best. We suspect that this is because the summaries are very short. More complex models, i.e., those with more topics, are not appropriate. The models perform differently at different levels of granularity because the amount of data supported at each time point depends on the chosen level. It is not necessarily the case that a finer grained model will contain enough data to provide a better predictive distribution.

### 4.3 Time Stamp Prediction

We can further use the cDTM for *time stamp prediction*, dating a document based on its content. To assess this task, we split each data set into 80% training and 20% testing sets. We predict the time stamp of each test document by finding its most likely location over the time line. We measure the error in terms of the same granularity at which the data are measured.

We investigated two approaches. The first is the *flat* approach. Each model of different granularity predicts as best it can. The second is the *hierarchical* approach. We use models of increasing granularity to "zoom in" on the prediction. For example, to predict the day, we first find the best month, then the best week within the month, and then the best day within the week.

We compute the average absolute error over the test data set. Figure 6 illustrates the results.

The hierarchical approach always performs better than or as well as the flat approach. The hour model in the AP data and day model in Election 08 perform worse. With the small data sets, a larger granularity is better. The reason may also lie in the parameter $v$. Currently it is shared among all models. In the future, we'd like to infer it from the data.

### 4.4 Example Topics

We provide some example topics by using the week model in the Election 08 data. We sample the topics every two months. Figure 7 shows one of the topics. At the beginning the election (year 2007), general issues were discussed more, such as "healthcare." As the competition went up (year 2008), the topics were more about candidates themselves and changing faster.

## 5 Conclusions

In this paper, we have developed the cDTM, using Brownian motion to model continuous-time topic evolution. The main advantage of the cDTM is that we can employ sparse variational inference for fast model comparison. We demonstrated the use of cDTM by measuring the predictive likelihood and time stamp prediction accuracy on two real-world data sets. In future work, we plan to explore the Ornstein-Uhlenbeck (OU) model [21], a generalization of Brownian model, that allows bounded variance.

***Acknowledgments***. We thank anonymous reviewers for their valuable comments. We would also like to thank Jordan Boyd-Graber and Jonathan Chang for many insightful discussions. David M. Blei is supported by ONR 175-6343, NSF CAREER 0745520, and grants from Google and Microsoft.

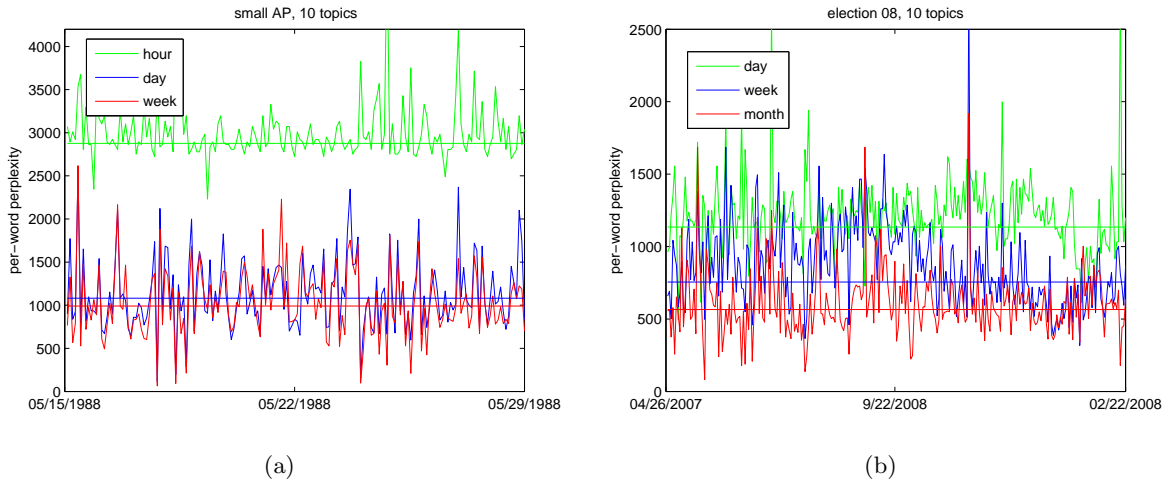

Figure 4: Per-word predictive perplexity comparison. The straight lines are the corresponding averaged per-word predictive perplexities. (a) AP data. The week model performs the best, but the day model is almost comparable. (b) Election 08 data. The month model performs the best.

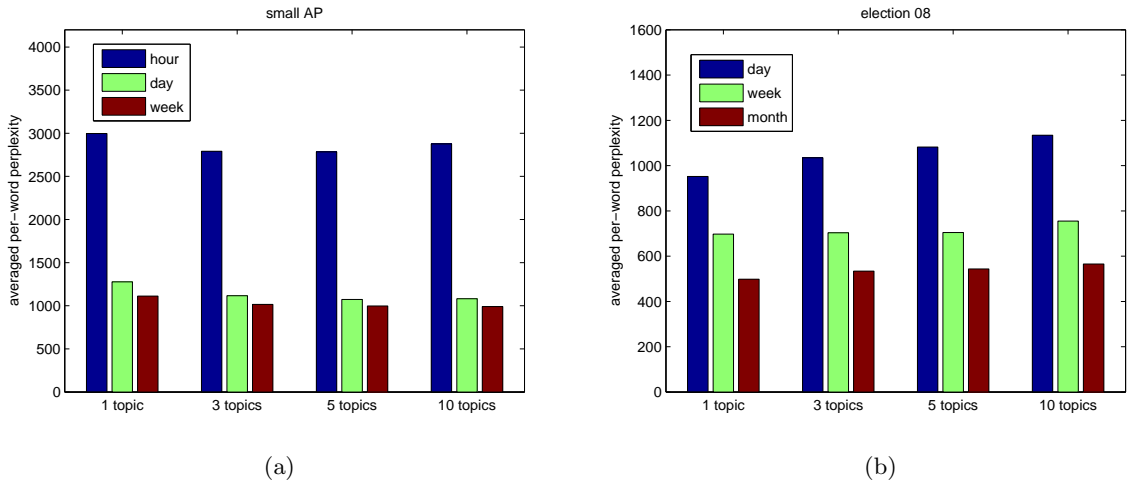

Figure 5: Averaged per-word predictive perplexity comparison. (a) AP data. The week model performs the best, but the day model is almost comparable. (b) Election 08 data. The month model performs the best.

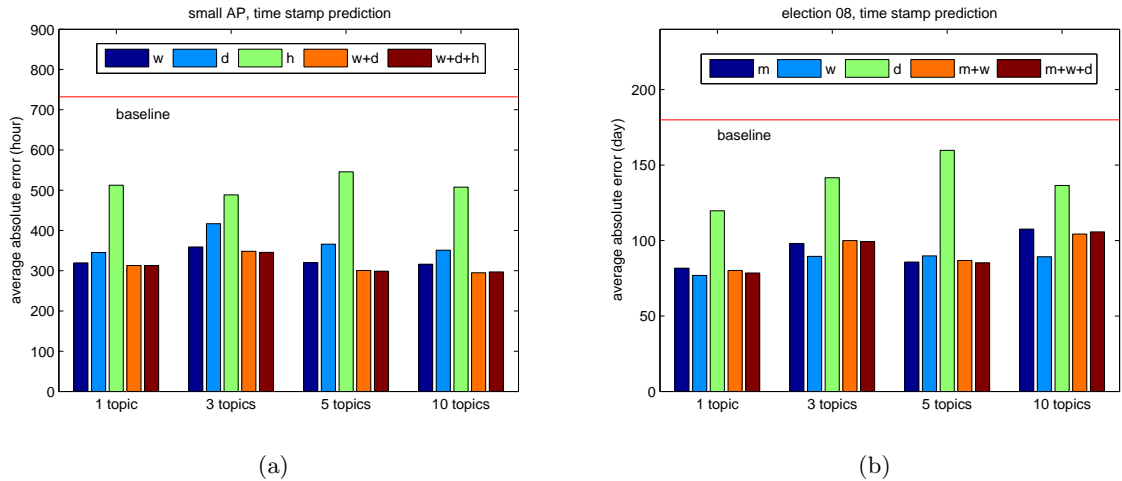

Figure 6: Time stamp prediction. 'm' stands for *flat* approach of 'month', 'w' for 'week', 'd' for 'day' and and 'h' for 'hour.' 'm+w' stands for the *hierarchical* approach of combining 'month' and 'week', and 'm+w+d', 'w+d', 'w+d+h' are similarly defined. The baseline is the expectation of the error by randomly assigning a time. (a) AP data. 5-topic and 10-topic models perform better than others, and the hierarchical approach always achieves the best performance. (b) Election 08 data. 1-topic model performs best due to the short documents. The hierarchical approach achieves comparable performances.

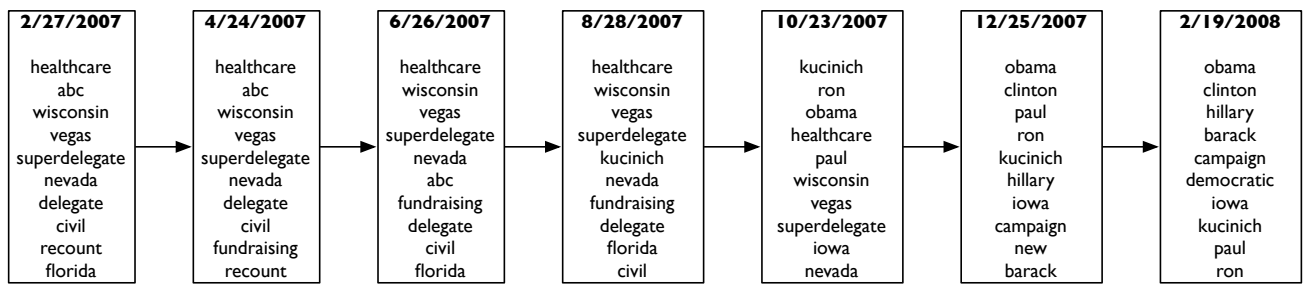

Figure 7: Examples from a 3-topic cDTM using the week model in the Election 08 data. In year 2007, the topics were more about general issues, while around year 2008, were more about candidates and changing faster.